\documentclass[aps,prd,reprint,superscriptaddress,nofootinbib,amsmath,amssymb,floatfix]{revtex4-2}
\usepackage[T1]{fontenc}
\usepackage{graphicx}
\usepackage{bm}
\usepackage{booktabs}
\usepackage{tikz}
\usepackage{verbatim}
\usetikzlibrary{arrows.meta,calc}
\usepackage[colorlinks=true,linkcolor=blue,citecolor=blue,urlcolor=blue]{hyperref}
\usepackage{amsmath,amssymb,amsthm}

\begin{document}

\title{Can Two-Point Correlations Determine Gravitational Waves from Sound Waves?}

\affiliation{School of Data Science and Artificial Intelligence, Wenzhou University of Technology, Wenzhou, China}

\author{Guangshang Chen}
\affiliation{
Key Laboratory of Particle Physics and Particle Irradiation (MOE), Institute of Frontier and Interdisciplinary Science,
Shandong University, Qingdao 266237, P. R. China
}

\author{Jianpeng Dai}
\email{Corresponding author, daijianpeng@wzut.edu.cn}
\affiliation{School of Data Science and Artificial Intelligence, Wenzhou University of Technology, Wenzhou, China}

\author{Yang Xiao}
\email{Corresponding author, xiaoyangphy@gmail.com}
\affiliation{
School of Physics, Henan Normal University,
Xinxiang 453007, P. R. China
}

\date{\today}

\begin{abstract}
Sound waves are expected to be a dominant source of gravitational waves (GWs) from cosmological first-order phase transitions. Their spectrum is commonly predicted using the fluid velocity two-point correlation function under the Gaussian approximation. However, since the GW source is quadratic in velocity, connected four-point correlations may also contribute to the GW spectrum. Through two analytical toy examples, we demonstrate that velocity two-point correlations alone does not always uniquely determine the GW spectrum. Moving beyond these idealized examples, we investigate a realistic acoustic source based on stochastic bubble nucleation and growth, proving that the connected contribution becomes subleading during subsequent linear acoustic propagation, although cosmic expansion limits this suppression. These results clarify when two-point reconstruction is reliable and how higher-order velocity correlations and the duration of acoustic emission affect GW predictions.
 
\end{abstract}

\maketitle

\section{Introduction}
First-order phase transitions in the early Universe can generate a stochastic GW background and provide a possible window into physics beyond the Standard Model~\cite{Athron:2023xlk}. During such transitions, bubbles of the new phase nucleate, expand, and eventually collide. Their expansion and collisions transfer energy to the surrounding plasma and generate large-scale fluid motion~\cite{Espinosa:2010hh}. The resulting sound waves can persist long after the bubble collisions and are expected to dominate GW production in a thermal plasma~\cite{Hindmarsh:2013xza,Hindmarsh:2015qta}. In particular, first-order phase transitions near the electroweak scale can generate GWs in the millihertz frequency range, which may be detectable by future space-based observatories such as LISA~\cite{amaro2017laser}, TianQin~\cite{luo2016tianqin}, and Taiji~\cite{ruan2020taiji}. The detection of such signals would open a new avenue for exploring particle physics and the early Universe~\cite{Caprini:2019egz,Kamionkowski:1993fg,Kosowsky:1992rz,Apreda:2001us,Grojean:2006bp,Randall:2006py,Schwaller:2015tja,Huang:2016odd,Bian:2017wfv,Chen:2017cyc}.

Considerable progress has been made in improving the predictions of acoustic GW production. Early numerical simulations characterized the acoustic GW spectrum~\cite{Hindmarsh:2017gnf} and motivated the development of the sound shell model (SSM), which linked its spectral features to the fluid shells surrounding expanding bubbles \cite{Hindmarsh:2016lnk}. The model was subsequently developed into a quantitative framework by combining hydrodynamic solutions for individual bubble profiles with the distributions of bubble lifetimes and different nucleation histories \cite{Hindmarsh:2019phv,Cai:2023guc}. This allowed the GW spectrum to be calculated directly from the parameters of the phase transition, rather than relying solely on phenomenological fits to numerical simulations. Further studies revisited the related time integrals and recovered the causal $k^3$ scaling in the infrared, showing that the previously predicted $k^9$ behavior does not apply in the strict low-frequency limit \cite{RoperPol:2023dzg,Sharma:2023mao}. Meanwhile, 
the effects of cosmic expansion and the finite acoustic lifetime were also investigated~\cite{Ellis:2020awk}, with the resulting suppression of the GW amplitude near the spectral peak characterized by the factor $\Upsilon$~\cite{Xiao:2024rsj, Guo:2020grp}.

Despite this progress, the statistical assumptions underlying acoustic GW predictions remain less well understood. In the SSM, the fluid velocity field is approximated as a linear superposition of sound waves generated by individual bubbles~\cite{Hindmarsh:2016lnk,Hindmarsh:2019phv}. If a sufficiently large number of independent contributions overlap, the central limit theorem suggests that the resulting velocity field may approach Gaussian statistics. This assumption allows Wick's theorem to reduce the four-point velocity correlations required for GW calculations to products of two-point functions. However, limited bubble nucleation and growth can generate nontrivial spatial and temporal correlations among bubble walls and nucleation sites~\cite{Megevand:2020klf,DeLuca:2021mlh,Pirvu:2021roq}. These correlations may be inherited by the fluid velocity field, raising questions about the validity of the Gaussian approximation. Moreover, nonlinear fluid interactions can invalidate the linear superposition of individual sound waves. Both effects may introduce extra contributions to the GW spectrum. Recent numerical simulations have found deviations from the Gaussian approximation in the fluid stress spectrum, while the resulting GW spectrum remains relatively well described by Gaussian-based predictions \cite{Correia:2025qif}. A subsequent analytical study showed that this approximation can predict a nonzero GW source before fluid-shell collisions, even though the exact contribution vanishes due to spherical symmetry \cite{Caprini:2026nnk}. These findings raise a fundamental question: under what conditions are velocity two-point correlations sufficient to determine the acoustic GW spectrum, and how does the contribution of higher-order correlations evolve during GW emission?

In this work, we investigate the role of non-Gaussian velocity correlations in acoustic GW production. We demonstrate that velocity two-point correlations do not, in general, uniquely determine the GW spectrum. Furthermore, we find that the contribution of higher-order correlations can become subleading as acoustic GW emission continues, although cosmic expansion limits this reduction. The remainder of this paper is organized as follows. In Sec.~\ref{sec:two_point_gw}, we introduce the theoretical framework and use two representative examples to illustrate the relationship between velocity two-point correlations and GW production. In Sec.~\ref{sec:real condition}, we investigate how the contribution of higher-order correlations to the GW signal evolves over time. Finally, we summarize our conclusions in Sec.~\ref{sec:conclusion}. Further details of the derivations are provided in the App.~\ref{sec:appendix}.

\section{Two-Point Correlations and Gravitational Wave Production}
\label{sec:two_point_gw}
GWs are sourced by the transverse-traceless (TT) component of the spatial energy-momentum tensor. On a flat background, the tensor perturbations $h_{ij}$ obey the linearized Einstein equation
\begin{equation}
 \ddot h_{ij}(\mathbf k,t)+k^2h_{ij}(\mathbf k,t)
 =16\pi G\Pi_{ij}(\mathbf k,t).
 \label{eq:sec2_tensor}
\end{equation}
Here $\Pi_{ij}=\Lambda_{ij,ab}T_{ab}$ denotes the TT source, with the projection operator $\Lambda_{ij,ab}=P_{ia}P_{jb}-P_{ij}P_{ab}/2$ and $P_{ij}=\delta_{ij}-\hat k_i\hat k_j$.

During a cosmological first-order phase transition, expanding bubbles induce fluid motions in the surrounding plasma, which in turn source GWs. In the acoustic regime, the plasma can be described as a perfect fluid with three-velocity $v_a$, whose spatial stress tensor is
\begin{equation}
    T_{ab}=w\gamma^2v_av_b+p_{\rm f}\delta_{ab},
\end{equation}
where $w$ and $p_{\rm f}$ are the fluid enthalpy density and pressure, $\gamma$ is the Lorentz factor. For weak acoustic perturbations, we approximate the enthalpy density by its background value, $w\simeq\bar w$, and take $\gamma\simeq1$. Since the isotropic pressure term vanishes under TT projection, the Fourier component of the TT anisotropic stress tensor at leading order is given by
\begin{equation}
\Pi_{ij}(\mathbf k,t)
=\bar w\Lambda_{ij,ab}
\int\frac{d^3q}{(2\pi)^3}
v_a(\mathbf q,t)v_b(\mathbf k-\mathbf q,t).
\label{eq:pi}
\end{equation}
The GW signal generated by this acoustic source can be characterized by its energy density and dimensionless energy-density spectrum,
\begin{equation}
\rho_{\rm GW}=\frac{\langle\dot h_{ij}\dot h_{ij}\rangle}{32\pi G},
\qquad
\Omega_{\rm GW}(k)=\frac{1}{\rho_{\rm tot}}
\frac{d\rho_{\rm GW}}{d\ln k}.
\end{equation}
Here the brackets denote an average over GW oscillations, and $\rho_{\rm tot}$ is the total background energy density. To relate the GW spectrum to the acoustic source, we solve Eq.~\eqref{eq:sec2_tensor} to express the tensor perturbations in terms of the TT anisotropic stress
\begin{equation}
\dot h_{ij}(\mathbf k,t)=16\pi G\int_{t_i}^{t_f}dt'
\cos[k(t-t')]\Pi_{ij}(\mathbf k,t'),
\label{eq}
\end{equation}
where $t_i$ and $t_f$ denote the beginning and end of the acoustic source activity, respectively. Since the GW energy density is quadratic in $\dot h_{ij}$, it is determined by the unequal-time correlator (UETC) of anisotropic stress, defined as~\cite{Hindmarsh:2019phv}
\begin{equation}
\langle\Pi_{ij}(\mathbf k,t_1)\Pi_{ij}^*(\mathbf k',t_2)\rangle
=(2\pi)^3\delta^3(\mathbf k-\mathbf k')U_\Pi(k;t_1,t_2).
\label{eq:uetc}
\end{equation}
Substituting Eq.~\eqref{eq:uetc} into Eq.~\eqref{eq} and averaging over GW oscillations yields
\begin{equation}
\Omega_{\rm GW}(k)=\frac{2Gk^3}{\pi\rho_{\rm tot}}
\int_{t_i}^{t_f} dt_1dt_2
\cos[k(t_1-t_2)]U_\Pi(k;t_1,t_2).
\label{eq:omega}
\end{equation}
The cosine factor accounts for the relative phases of GWs emitted at different times.

Since the anisotropic stress is quadratic in the plasma velocity, its UETC involves the velocity four-point correlation function. In the SSM, the velocity field $v_a$ is approximated as a linear superposition of sound shells generated by individual expanding bubbles~\cite{Hindmarsh:2016lnk,Hindmarsh:2019phv}. When many approximately independent sound shells contribute to the velocity field, the central limit theorem motivates treating it as approximately Gaussian. For a statistically homogeneous, isotropic, and irrotational acoustic velocity field, the corresponding UETC takes the form~\cite{Hindmarsh:2019phv}
\begin{equation}
\langle v_i(\mathbf q,t_1)v_j^*(\mathbf q',t_2)\rangle
=(2\pi)^3\delta^3(\mathbf q-\mathbf q')
\hat q_i\hat q_jG(q;t_1,t_2).
\end{equation}
The four-point function entering the stress UETC can therefore be decomposed via Wick's theorem:
\begin{align}
\langle v_1v_2v_3v_4\rangle_{\rm wick}
&=\langle v_1v_2\rangle\langle v_3v_4\rangle
+\langle v_1v_3\rangle\langle v_2v_4\rangle\nonumber \\
&+\langle v_1v_4\rangle\langle v_2v_3\rangle.
\end{align}
However, the Gaussian approximation is not guaranteed in realistic phase transitions. Nonlinear fluid evolution can invalidate the linear superposition of individual sound shells, while a finite number of bubble nucleation events and subsequent bubble growth can generate spatial correlations among bubbles that are inherited by the resulting fluid motions. Consequently, the velocity field may retain non-Gaussian correlations, giving rise to an additional connected four-point function
\begin{align}\label{eq:4pt_resolved}
\langle v_1v_2v_3v_4\rangle = 
\langle v_1v_2v_3v_4\rangle_{\rm wick} + \langle v_1v_2v_3v_4\rangle_c.
\end{align}
This connected correlation contributes an additional term to the anisotropic-stress UETC, $U_\Pi=U_\Pi^{\rm W}+U_\Pi^c$, where $U_\Pi^{\rm W}$ is the Wick contribution and $U_\Pi^c$ is the TT-projected connected contribution. The latter can modify the predicted GW spectrum, indicating that the velocity two-point function alone does not necessarily determine the acoustic GW spectrum. To illustrate this point, the following toy examples show both possibilities.
\begin{figure*}[t]
 \centering
 \includegraphics[width=\textwidth]{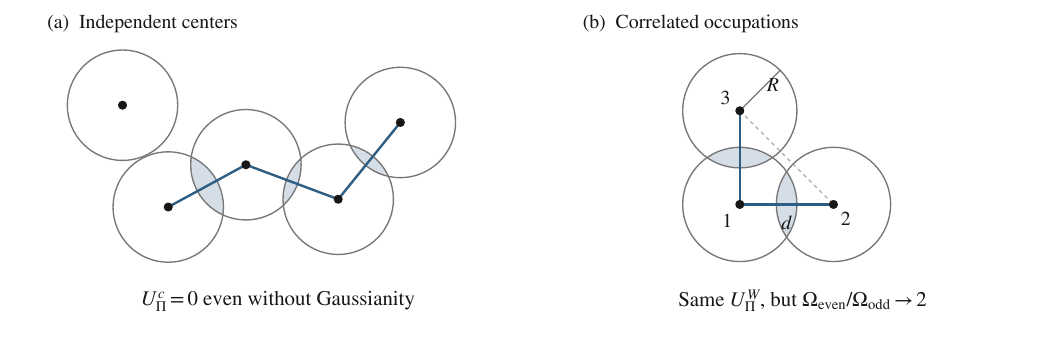}
 \caption{Sound shell configurations in the two toy examples. Shaded regions indicate shell overlaps. (a) Poisson-distributed shells: the TT self-stress of each spherical shell vanishes, while overlaps between distinct shells can source GWs. (b) Three-site configuration: only pairs $(1,2)$ and $(1,3)$ overlap. The even and odd ensembles have identical velocity two-point correlations but different GW spectra.}
 \label{fig:sec2_example_pairs}
\end{figure*}

\subsection{Toy example I: Exact two-point reconstruction for Poisson sound shells}

We first consider a collection of spherically symmetric sound shells whose contributions to the plasma velocity add linearly,
\begin{equation}
v_i(\mathbf x,t)=\sum_a u_i(\mathbf x-\mathbf X_a,t;m_a)=\sum_a u_i^{m_a}(\mathbf x-\mathbf X_a,t),
\label{eq:toy1_velocity}
\end{equation}
where $\mathbf X_a$ denotes the center of the $a$-th shell. We assume that the centers follow a homogeneous Poisson process with number density $n$, while the independent marks $m_a$ characterize the radial profiles of individual shells, accounting for their different formation times. For a finite number of contributing sound shells, the velocity field need not be Gaussian. Nevertheless, as shown below, its connected four-point contribution to the GW spectrum vanishes.

For an individual sound shell, spherical symmetry constrains the Fourier transform of its quadratic velocity stress to the form
\begin{equation}
\mathcal F[u_a^m u_b^m]
=A_m\delta_{ab}+B_m\hat k_a\hat k_b,
\label{eq:toy1_selfstress}
\end{equation}
where $A_m$ and $B_m$ are scalar functions of $k$ and $t$. Both terms vanish under TT projection, since $\Lambda_{ij,ab}\delta_{ab}=\Lambda_{ij,ab}\hat k_a\hat k_b=0$. Defining the TT self-stress of a single shell as $S_{ij}^m=\Lambda_{ij,ab}\mathcal F[u_a^m u_b^m]$, we therefore obtain $S_{ij}^m=0$. However, when two shells overlap, their cross terms need not be spherically symmetric and can generate a nonzero TT stress. The GW source therefore arises from the overlaps of distinct sound shells, as illustrated in Fig.~\ref{fig:sec2_example_pairs}.

To evaluate the connected velocity four-point function, we consider the velocity field in Eq.~\eqref{eq:toy1_velocity} at four spacetime points $(\mathbf x_\alpha,t_\alpha)$, with $\alpha=1,\ldots,4$. Introducing auxiliary variables $J_\alpha$, the generating function for a fixed number $N$ of sound shells can be written as
\begin{align}
&\left\langle e^{\sum_{\alpha=1}^4J_\alpha v_{i_\alpha}(\mathbf x_\alpha,t_\alpha)}\mid N\right\rangle
\nonumber\\
&=\left\langle
\prod_{a=1}^{N}
e^{\sum_{\alpha=1}^4J_\alpha u_{i_\alpha}^{m_a}(\mathbf x_\alpha-\mathbf X_a,t_\alpha)}
\Bigm|N
\right\rangle
\nonumber\\
&=\prod_{a=1}^{N}
\left\langle
e^{\sum_{\alpha=1}^4J_\alpha u_{i_\alpha}^{m_a}(\mathbf x_\alpha-\mathbf X_a,t_\alpha)}
\right\rangle_{\mathbf X_a,m_a}
\nonumber\\
&=[g(\mathbf J)]^N.
\label{eq:toy1_fixedN}
\end{align}
The second equality follows from the statistical independence of different sound shells. Since their centers are uniformly distributed and their marks are drawn from the same probability distribution, each factor has the common value
\begin{equation}
g(\mathbf J)
=\frac{1}{V_{\rm tot}}
\int_{V_{\rm tot}}d^3X
\left\langle
e^{\sum_{\alpha=1}^{4}J_\alpha u_{i_\alpha}^m(\mathbf x_\alpha-\mathbf X,t_\alpha)}
\right\rangle_m,
\label{eq:toy1_single_generator}
\end{equation}
where the spatial integral averages over the center of a single shell, while $\langle\cdots\rangle_m$ denotes the average over its mark.

The shell number $N$ follows a Poisson distribution with mean $nV_{\rm tot}$. Averaging over $N$ therefore gives
\begin{align}
&\left\langle e^{\sum_{\alpha=1}^4J_\alpha v_{i_\alpha}(\mathbf x_\alpha,t_\alpha)}\right\rangle
\nonumber\\
&=\sum_{N=0}^{\infty}
\frac{e^{-nV_{\rm tot}}(nV_{\rm tot})^N}{N!}
[g(\mathbf J)]^N
\nonumber\\
&=\exp\left[nV_{\rm tot}(g(\mathbf J)-1)\right].
\label{eq:toy1_poisson_sum}
\end{align}
In the infinite-volume limit, taking the logarithm and substituting $g(\mathbf J)$, we obtain the cumulant-generating function
\begin{align}
&\ln\left\langle e^{\sum_{\alpha=1}^4J_\alpha v_{i_\alpha}(\mathbf x_\alpha,t_\alpha)}\right\rangle
\nonumber\\
&=n\int d^3X
\left[
\left\langle
e^{\sum_{\alpha=1}^4J_\alpha u_{i_\alpha}^m(\mathbf x_\alpha-\mathbf X,t_\alpha)}
\right\rangle_m-1
\right],
\label{eq:toy1_poisson_generator}
\end{align}
where the infinite-volume limit is understood. Taking one derivative with respect to each $J_\alpha$ and setting all $J_\alpha=0$ yields
\begin{align}
&\left\langle
\prod_{\alpha=1}^{4}
v_{i_\alpha}(\mathbf x_\alpha,t_\alpha)
\right\rangle_c
\nonumber\\
&=n\int d^3X
\left\langle
\prod_{\alpha=1}^{4}
u_{i_\alpha}^m(\mathbf x_\alpha-\mathbf X,t_\alpha)
\right\rangle_m.
\label{eq:toy1_poisson_cumulant}
\end{align}
Thus, all four velocity factors in the connected contribution originate from the same shell, sharing a common center $\mathbf X$ and mark $m$. Its contribution to the stress UETC is therefore
\begin{equation}
U_\Pi^c(k;t_1,t_2)
=n\bar w^2
\left\langle
S_{ij}^m(\mathbf k,t_1)
S_{ij}^{m*}(\mathbf k,t_2)
\right\rangle_m,
\label{eq:example_poisson_connected}
\end{equation}
where the brackets denote an average over shell marks. Since the TT self-stress of every spherical shell vanishes, $S_{ij}^m=0$, the connected contribution satisfies $U_\Pi^c=0$ at all unequal times. The full stress UETC is therefore exactly reconstructed from the velocity two-point function, $U_\Pi=U_\Pi^{\rm W}$.

\subsection{Toy example II: Identical velocity correlations but different GW spectra}

The preceding example shows that Gaussianity is not necessary for exact two-point reconstruction. However, this does not imply that such reconstruction is always valid. Consider three candidate sites for sound shell centers at
$\mathbf X_1=(0,0,0)$, $\mathbf X_2=(d,0,0)$, and $\mathbf X_3=(0,d,0)$. Each site has an occupation probability of $1/2$ and, when occupied, carries an identical spherically symmetric sound shell whose radial velocity profile vanishes outside a radius $R$. We choose
\begin{equation}
d<2R<\sqrt{2}d,
\end{equation}
so that sound shells 1 and 2, as well as sound shells 1 and 3, can overlap, whereas sound shells 2 and 3 cannot, as illustrated in Fig.~\ref{fig:sec2_example_pairs}. We assign a binary occupation variable $\eta_a\in \{0,1\}$ to each site, indicating whether the corresponding sound shell is present. Following the linear superposition assumed in the SSM, the velocity field of a cluster is given by
\begin{equation}
v_i(\mathbf x,t)
=\sum_{a=1}^{3}\eta_a u_i(\mathbf x-\mathbf X_a,t),
\label{eq:toy2_velocity}
\end{equation}
where $\eta_a=1$ indicates that the corresponding shell is present. 

We compare two occupation distributions. In the even ensemble, the configurations $000$, $011$, $101$, and $110$ each occur with probability $1/4$, while in the odd ensemble, $001$, $010$, $100$, and $111$ have the same probabilities. Both distributions satisfy
\begin{equation}
\langle\eta_a\rangle=\frac12,\qquad
\langle\eta_a\eta_b\rangle=\frac14
\quad(a\ne b).
\label{eq:toy2_pair_moments}
\end{equation}
The corresponding velocity two-point correlation function is given by
\begin{equation}
\langle v_i v_j\rangle
=\sum_{a,b=1}^{3}
\langle\eta_a\eta_b\rangle_\eta
u_i^{(a)}u_j^{(b)}.
\label{eq:toy2_velocity_uetc}
\end{equation}
Here $\langle\cdots\rangle_\eta$ denotes the average over occupation variables. Since $\langle\eta_a\eta_b\rangle_\eta=1/2$ for $a=b$ and $1/4$ for $a\ne b$ in both ensembles, the even and odd ensembles have identical unequal-time velocity correlation functions.

To examine whether the two ensembles also produce the same GW spectrum, we consider the TT anisotropic stress. As shown in the preceding example, the TT self-stress of an individual spherical shell vanishes, leaving only cross terms between distinct shells. Defining the TT stress generated by a sound shell pair as
\begin{equation}
\Pi_{ij}^{ab}(\mathbf k,t)
=\bar w\Lambda_{ij,lm}
\mathcal F[u_l^{(a)}u_m^{(b)}
+u_l^{(b)}u_m^{(a)}],
\label{eq:toy2_pair_stress}
\end{equation}
the total TT stress in our three-site configuration reduces to
\begin{equation}
\Pi_{ij}(\mathbf k,t)
=\eta_1\eta_2\Pi_{ij}^{12}(\mathbf k,t)
+\eta_1\eta_3\Pi_{ij}^{13}(\mathbf k,t),
\label{eq:toy2_total_stress}
\end{equation}
where the contribution from the pair $(2,3)$ vanishes because the two sound shells do not overlap. Substituting this expression into the stress UETC gives
\begin{align}
&\langle\Pi_{ij}(t_1)\Pi_{ij}^*(t_2)\rangle_\eta
\nonumber\\
&=b\left[
\Pi_{ij}^{12}(t_1)\Pi_{ij}^{12*}(t_2)
+\Pi_{ij}^{13}(t_1)\Pi_{ij}^{13*}(t_2)
\right]
\nonumber\\
&\quad+t\left[
\Pi_{ij}^{12}(t_1)\Pi_{ij}^{13*}(t_2)
+\Pi_{ij}^{13}(t_1)\Pi_{ij}^{12*}(t_2)
\right],
\label{eq:toy2_stress_uetc}
\end{align}
where $b=\langle\eta_1\eta_2\rangle_\eta=\langle\eta_1\eta_3\rangle_\eta=1/4$, and $t=\langle\eta_1\eta_2\eta_3\rangle_\eta$. The latter arises because $\eta_a$ is a binary variable, giving $\eta_1^2\eta_2\eta_3=\eta_1\eta_2\eta_3$. Notably, unlike $b$, the three-site occupation moment differs between the two ensembles: $t=0$ for the even ensemble and $t=1/4$ for the odd ensemble.

The corresponding GW spectra follow from Eq.~\eqref{eq:omega}, which involves a double-time integral of the stress UETC with the kernel $\cos[k(t_1-t_2)]$. Using $\cos[k(t_1-t_2)]=[e^{ik(t_1-t_2)}+e^{-ik(t_1-t_2)}]/2$, the double-time integrals can be expressed as products of single-time integrals of the form
\begin{equation}
B_{ij}^{ab}(\mathbf k)
=\int_{t_i}^{t_f}dt~
e^{ikt}\Pi_{ij}^{ab}(\mathbf k,t).
\end{equation}
After averaging over propagation directions, the resulting GW spectrum can be expressed, up to common normalization factors, as
\begin{align}
\Omega_{\rm GW}(k)\propto k^3\Big[
&b\,\overline{B_{ij}^{12}B_{ij}^{12*}+B_{ij}^{13}B_{ij}^{13*}}
\nonumber\\
&+2t\,\overline{\operatorname{Re}
\left(B_{ij}^{12}B_{ij}^{13*}\right)}
\Big],
\label{eq:toy2_gw_spectrum}
\end{align}
where the overbar denotes an average over GW propagation directions, and repeated spatial indices are summed.

In the low-frequency limit, the two contributing sound shell pairs, $(1,2)$ and $(1,3)$, have the same separation $d$, along the $x$ and $y$ directions, respectively. Their TT amplitudes can therefore be written as
\begin{equation}
B_{ij}^{12}=B_0\Lambda_{ij,lm}\hat x_l\hat x_m,
\qquad
B_{ij}^{13}=B_0\Lambda_{ij,lm}\hat y_l\hat y_m,
\label{eq:toy2_lowk_amplitudes}
\end{equation}
where $B_0$ is a common nonzero scalar coefficient determined by the sound shell profile and the source duration. For two arbitrary unit vectors $\hat{\mathbf a}$ and $\hat{\mathbf b}$, the isotropic TT average satisfies
\begin{equation}
\overline{
\Lambda_{ij,lm}\hat a_l\hat a_m
\Lambda_{ij,pq}\hat b_p\hat b_q
}
=\frac25\left[
(\hat{\mathbf a}\cdot\hat{\mathbf b})^2-\frac13
\right].
\label{eq:toy2_tt_average}
\end{equation}
For the two perpendicular directions, the contributions from the pairs $(1,2)$ and $(1,3)$ each give $4/15$, while their cross term gives $-2/15$. Substituting these results into Eq.~\eqref{eq:toy2_gw_spectrum}, we obtain
\begin{equation}
\Omega_{\rm GW}(k)
\propto k^3\frac{4|B_0|^2}{15}(2b-t),
\qquad k\to0^+.
\label{eq:toy2_ir_gw}
\end{equation}
Using $b=1/4$, $t_{\rm even}=0$, and $t_{\rm odd}=1/4$, we find
\begin{equation}
\lim_{k\to0^+}
\frac{\Omega_{\rm GW}^{\rm even}(k)}
{\Omega_{\rm GW}^{\rm odd}(k)}
=2.
\label{eq:toy2_gw_ratio}
\end{equation}

Thus, the two ensembles have identical velocity two-point correlations but produce different GW spectra. However, the mean velocity of a fixed three-site configuration is generally nonzero, so the zero-mean Wick decomposition cannot be directly applied. To address this, we consider many copies of the three-site configuration randomly distributed in spacetime, with independent uniform orientations. Their candidate centers are drawn from a homogeneous Poisson process, and those with other candidates within a prescribed exclusion region are removed. The exclusion region is chosen to preserve spatial isotropy and prevent sound shells from different configurations from overlapping during their active periods.

The resulting velocity field is statistically homogeneous and isotropic, and hence has zero ensemble mean, $\langle v_i\rangle=0$. The even and odd ensembles also retain identical velocity UETCs, since their single-site and pairwise occupation probabilities remain unchanged. We label each three-site configuration by $A$ and denote its TT stress by $\Pi_{ij}^{(A)}$. Since shells belonging to different configurations do not overlap, the total TT stress can be written as $\Pi_{ij}=\sum_A\Pi_{ij}^{(A)}$.

For each configuration $A$, averaging over its uniformly random orientation gives
\begin{equation}
\left\langle\Pi_{ij}^{(A)}(\mathbf k,t)\right\rangle_{\mathcal R_A}
=\Lambda_{ij,lm}
\left[C_A\delta_{lm}+D_A\hat k_l\hat k_m\right]=0,
\label{eq:toy2_orientation_average}
\end{equation}
where $\langle\cdots\rangle_{\mathcal R_A}$ denotes the orientation average, and $C_A$ and $D_A$ are scalar coefficients. The result follows because both tensor structures vanish under TT projection. Since the orientations of different configurations are independent, their cross terms factorize as
\begin{align}
&\left\langle
\Pi_{ij}^{(A)}(\mathbf k,t_1)
\Pi_{ij}^{(B)*}(\mathbf k,t_2)
\right\rangle_{\mathcal R_A,\mathcal R_B}
\nonumber\\
&=
\left\langle\Pi_{ij}^{(A)}(\mathbf k,t_1)\right\rangle_{\mathcal R_A}
\left\langle\Pi_{ij}^{(B)*}(\mathbf k,t_2)\right\rangle_{\mathcal R_B}
=0,\qquad A\ne B.
\label{eq:toy2_cross_configuration}
\end{align}
Here $\mathcal R_A$ and $\mathcal R_B$ denote independent orientations of configurations $A$ and $B$, with their centers and occupation states fixed. Consequently, the contributions from different configurations add in the ensemble-averaged GW spectrum, while the interference between shell pairs within each configuration is preserved. The factor-of-two difference between the even and odd ensembles therefore remains unchanged. Since both ensembles have zero mean velocity and identical velocity UETCs, their Wick contributions are identical. The difference in their GW spectra must therefore arise from the TT-projected connected velocity four-point correlation. 

\section{Time Evolution of Connected Correlation Effects}
\label{sec:real condition}


The toy examples in Sec.~\ref{sec:two_point_gw} demonstrate that the fluid velocity two-point correlation function does not always uniquely determine the acoustic GW spectrum. We now investigate the contribution of connected four-point velocity correlations to the GW spectrum in a more physically motivated scenario.

\subsection{Sound waves from expanding bubbles}
During a first-order phase transition, bubbles of the new phase nucleate and expand, which in turn induces fluid motion in the surrounding plasma. Typically, bubble nucleation can be modeled as a Poisson process, with a nucleation rate per unit volume and unit time given by $\Gamma(u)=\Gamma_0 e^{\beta u}$, where $u=t-t_{\rm nuc}$ and $t_{\rm nuc}$ denotes the end of the nucleation stage. As these bubbles expand, they convert the surrounding regions to the new phase. To describe this process, we introduce a boolean function $\chi(\bm{x},t)$, which equals unity if the position $\bm{x}$ has converted to the new phase by time $t$, and zero otherwise. For a spacetime point $z_a=(\bm{x}_a,t_a)$, we denote $\chi_a=\chi(\bm{x}_a,t_a)$ and define ${\cal C}_a$ as the spacetime region of earlier nucleation events whose expanding bubbles can reach $z_a$. Since a point remains in the old phase only if no such nucleation event occurs, the Poisson probability of no events, together with the inclusion–exclusion principle, yields the joint conversion moments \cite{Megevand:2020klf}:
\begin{equation}
    \bigg\langle\prod_{a=1}^m\chi_a\bigg\rangle = \sum_{A\subseteq \{1,2,\dots,m\}}(-1)^{|A|}e^{-\Lambda(\cup_{a\in A}{\cal C}_a)}.
\end{equation}
Here $|A|$ denotes the number of elements in the subset $A$, and $\Lambda(D)$ represents the mean number of nucleation events in a spacetime region $D$, defined as
\begin{equation}
\Lambda(D) = \int_D {\rm d}^3{\bm x}\,{\rm d}s\,\Gamma(s).
\end{equation}

Having described the stochastic nature of the phase transition, we now consider the resulting fluid motion. Specifically, the energy density and pressure perturbations can be decomposed into thermal and phase-transition contributions as $\delta e=\delta e_{\rm th}+\Delta e\chi$ and $\delta p=c_s^2\delta e_{\rm th}+\Delta p\chi$, respectively. Substituting these expressions into the fluid energy-momentum tensor and imposing energy-momentum conservation, we obtain the following linearized wave equation for the fluid velocity $\bm v$ \cite{Jinno:2022mie,Jinno:2024nwb}:
\begin{equation}\label{eq:conversion_driving}
    (\partial_t^2 - c_s^2\nabla^2)\bm v = -\alpha_\chi\nabla\dot\chi,
\end{equation}
where $\alpha_\chi$ characterizes the strength of the phase transition and is given by~\cite{Giese:2020rtr}
\begin{equation}
\alpha_\chi=\frac{\Delta p-c_s^2\Delta e}{w_0}
=\frac{\epsilon_{\rm vac}}{\rho_{\rm rad}}.
\end{equation}
Here $w_0$ is the background enthalpy density. This equation shows that the evolution of the fluid velocity depends on the phase transition history encoded in $\chi$.



Although the phase transition occurs over a finite time interval, the resulting sound waves can continue to propagate and source GWs after the transition. To study the accumulated GW emission during this subsequent evolution, we introduce a smooth time window $\tau(u)$ to describe the finite duration effect, together with a real, normalized, isotropic spatial smoothing kernel $K_\ell$ with characteristic scale $\ell$. Solving Eq.~\eqref{eq:conversion_driving} in Fourier space, the fluid velocity can be expressed as
\begin{equation}
    \hat{\bm v}(\bm q,s) = \hat K_\ell(\bm q)\int{\rm d}u\,{\cal K}(\bm q;u,s)\hat\chi(\bm q,u),
\end{equation}
where $\omega_q=c_s q$ and
\begin{equation}\label{eq:pt2v}
    {\cal K}(\bm q;u,s) = \frac{\alpha_\chi\bm q}{2\omega_q}\sum_{\sigma=\pm} [\sigma \tau'(u)-i\omega_q \tau(u)]e^{i\sigma\omega_{q}(s-u)}.
\end{equation}
After the driving stage ends at $s=0$, the sound waves propagate freely and continue to generate GWs over the interval $0\leq s\leq T_{\rm obs}$.

\subsection{Different accumulation of Wick and connected contributions}

We first investigate the spatial correlations induced by bubble nucleation and growth. Consider four points $\bm{x}_i$ ($i=1,2,3,4$) at $u=0$, with a maximum separation $D=\max_{i,j}|\bm{x}_i-\bm{x}_j|$. Bubbles nucleated at earlier times have more time to expand and can therefore influence other regions. To quantify this effect, we take the bubble wall velocity $v_w$ to be constant and introduce a truncated indicator function $\chi_i^R$ that excludes nucleation events before $u=-R/v_w$. Consequently, any remaining nucleation event capable of reaching $\bm{x}_i$ must lie within a distance $R$ of that point.


Excluding early nucleation events may change the value of $\chi_i$ if an excluded bubble would otherwise have reached $\bm{x}_i$. However, the probability of such an event is exponentially suppressed and satisfies
\begin{equation}
    {\rm Pr}(\chi_i^R\neq\chi_i)
    \leq
    \frac{8\pi v_w^3\Gamma_0}{\beta^4}
    e^{-\frac{\beta R}{v_w}}
    Q_3\left(\frac{\beta R}{v_w}\right),
\end{equation}
where $Q_3(z)=1+z+\frac12z^2+\frac16z^3$.
The spatial restriction also allows us to establish the independence of sufficiently separated regions. In particular, two truncated indicators are statistically independent if their corresponding regions do not overlap. If all four regions formed a connected cluster, the two most distant points could be linked by at most three steps, each of length no greater than $2R$.
However, choosing $R=D/12$ makes this condition impossible. In this case, the four truncated indicators can be divided into two independent nonempty groups, yielding a vanishing connected four-point correlation. The bound on the truncation error, together with the vanishing connected correlation of the truncated indicators, gives
\begin{equation}
    |\langle\chi_1\chi_2\chi_3\chi_4\rangle_c|
    \leq 104\Lambda_f e^{-\frac{\beta D}{12v_w}}
    Q_3\left(\frac{\beta D}{12v_w}\right).
\end{equation}
This result shows that the connected velocity four-point correlation decays exponentially with spatial separation.

We now consider how these correlations affect GW production. As discussed in Sec.~\ref{sec:two_point_gw}, the GW spectrum contains both Wick and connected contributions from the velocity four-point correlation function. Accordingly, we can write
\begin{equation}
    \Omega_{\rm GW}(k)\propto W_f(k)+C_f(k),
\end{equation}
where $W_f$ and $C_f$ denote the two contributions, respectively. The connected contribution accumulated during subsequent acoustic propagation can be expressed as
\begin{equation}
    C_f(k)=
    \int {\rm d}\tilde q\,{\rm d}\tilde q'\,
    H_C(\tilde q,\tilde q';k)
    F_f(\tilde q)F_f^*(\tilde q'),
    \label{eq:gw_connected_contribution}
\end{equation}
where $H_C$ contains the TT-projected connected four-point correlations, including the momentum integrations and sums over acoustic modes. A window function $f(s)=1$ within $0\le s\le T_{\rm obs}$ and zero otherwizse is introduced to ensure the observed GW emission duration, with $F_f$ its Fourier correspondence.
Detuning $\tilde q=k-\sigma\omega_q-\tau\omega_{|\bm k-\bm q|}$ ($\sigma,\tau=\pm1$) represents the frequency difference between an acoustic-mode pair and the GW mode, while $\tilde q'$ corresponds to the second stress factor. The double integral then includes all possible combinations of these frequency differences contributing to the GW spectrum.

The exponential spatial decay derived above, together with the finite driving duration and spatial smoothing $K_\ell$, yields the bound
\begin{equation}
    |H_C(\tilde q,\tilde q';k)|
    \leq
    \frac{Q_C(k)}
    {[1+(\tilde q/\beta)^2][1+(\tilde q'/\beta)^2]},
\end{equation}
where $Q_C(k)$ is finite and independent of the observation time. As a result, acoustic modes far from resonance undergo phase cancellation, which limits the accumulation of the connected contribution:
\begin{equation}
    |C_f(k)|\leq Q_C(k)J^2(E),
\end{equation}
where $J(E)=2\sqrt{\pi}+2\ln[1+(\beta E)^2]$, $\beta$ is the inverse duration of the phase transition and $E(T_{\rm obs})=\int_0^{T_{\rm obs}}{\rm d}s\,f^2(s)$ represents the effective exposure
of GW emission. In a flat spacetime, we have $E=T_{\rm obs}$. The GWs generated during the sound-wave generation stage contribute a term independent of $E$, while their interference with subsequent acoustic emission grows at most as $J(E)$. Therefore, the complete connected contribution $C_f(k)$ grows at most logarithmically squared with $E$.

In contrast, the Wick contribution has a different frequency structure. Its pairwise contractions match the acoustic momenta between the two stress factors, leading to
\begin{equation}
    H_W^{\rm res}(\tilde q,\tilde q';k)
    =Q_W(\tilde q;k)\delta_{\rm D}(\tilde q-\tilde q').
\end{equation}
This frequency matching allows the Wick contribution to accumulate linearly with the observation time. For $Q_W(0;k)>0$, we obtain
\begin{equation}
    W_f(k)=2\pi Q_W(0;k)E+o(E).
\end{equation}
Consequently, the relative connected contribution satisfies
\begin{equation}
    \frac{|C_f(k)|}{W_f(k)}
    =O\!\left[
    \frac{\ln^2(1+\beta E)}{\beta E}
    \right],
    \qquad E\to\infty.
\end{equation}
Therefore, the connected contribution becomes subleading during prolonged acoustic propagation because it lacks the frequency matching that allows the Wick contribution to accumulate linearly.

\subsection{The available acoustic exposure in the expanded universe}


The suppression of the connected contribution found above requires a sufficiently large value of $\beta E$. In flat spacetime, $E=T_{\rm obs}$ can grow with the duration of acoustic propagation. In an expanding Universe, however, cosmic expansion reduces the contribution of the source at later times and limits the growth of $E$. We therefore examine how cosmic expansion affects the relative contributions of the Wick and connected terms.

\begin{figure}[t]
 \centering
 \includegraphics[width=0.49\textwidth]{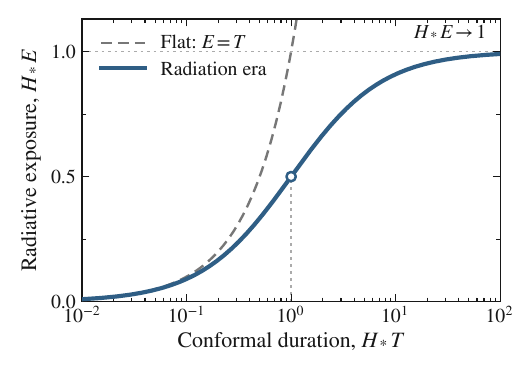}
 \caption{The time integral $E$ as a function of the conformal propagation time $T_{\rm obs}$. The solid curve shows the radiation-era result, $H_*E=H_*T_{\rm obs}/(1+H_*T_{\rm obs})$, while the dashed curve shows the flat-spacetime result, $E=T_{\rm obs}$.}
 \label{fig:sec3_compare}
\end{figure}

For electroweak scale phase transitions, the Universe is generally radiation dominated. Since bubble nucleation and collisions occur over a relatively short timescale, cosmic expansion can usually be neglected during these stages. Its effects should, however, be taken into account during subsequent acoustic propagation and GW production. We describe the expansion during this later stage using the conformal time $\eta$, measured from the end of sound wave generation. With the scale factor normalized to unity at $\eta=0$, a radiation-dominated background gives $a(\eta)=1+\eta H_*$. During radiation domination, the fluid enthalpy density scales as $a^{-4}$, while the amplitudes of linear acoustic velocity perturbations remain approximately constant. Consequently, the expansion introduces an additional factor $a^{-2}$ into the time integral for the accumulated GW power. The effective duration $E$ therefore becomes
\begin{equation}
E(T_{\rm obs}) =
\int_0^{T_{\rm obs}}{\rm d}\eta
\frac{f^2(\eta)}{a^2(\eta)}
= \frac{T_{\rm obs}}{1+H_*T_{\rm obs}},
\label{eq:E_expanding}
\end{equation}
where we have taken $f(\eta)=1$ during the propagation interval. Unlike the flat spacetime case, $E$ no longer grows indefinitely but approaches $1/H_*$ as $T_{\rm obs}\to\infty$, as shown in Fig.~\ref{fig:sec3_compare}.

On the other hand, the duration of the free-acoustic stage is also limited by nonlinear fluid effects. As sound waves propagate, nonlinear terms cause them to steepen and eventually lead to turbulence. Let $L_*$ denote the characteristic comoving scale of the fluid flow, and $U$ the rms fluid velocity. The characteristic time for the onset of nonlinear effects can be estimated as
\begin{equation}
    T_{\rm NL}\sim\frac{L_*}{U}.
\end{equation}

Nonlinear fluid effects generally develop long before the present day. Equation~\eqref{eq:E_expanding} then gives the effective acoustic exposure as 
\begin{equation}
    E
    =\frac{T_{\rm NL}}{1+H_*T_{\rm NL}}
    \lesssim
    \left(H_*+\frac{U}{L_*}\right)^{-1}.
\end{equation}
Consequently, in realistic scenarios, the connected four-point contribution to the GW spectrum in Eq.~\eqref{eq:gw_connected_contribution} need not be negligible. This implies that the GW spectrum may not be fully determined by the velocity two-point correlation function alone. 


\section{Conclusion}
\label{sec:conclusion}

In this work, we investigated the role of higher-order velocity correlations in acoustic GW production. Using two analytical toy examples, we demonstrated that the velocity two-point correlation function fails to uniquely fix the GW spectrum. In a more realistic phase transition, the finite duration of bubble nucleation and the low nucleation rate at early times limit the accumulation of connected correlations, thereby suppressing their contribution relative to the Wick contribution. Despite the dilution, the acoustic Non-Gaussianity still remain quantitatively relevant in the full GW spectrum due to the finite exposure imposed by the cosmological expansion and nonlinear transition.

Our work highlights the role of source evolution in shaping in determining the statistical features of the primordial GW background. However, the exact quantitative impact is model-dependent and a complete assessment must also account for subsequent non-linear phase. Non-linear fuild dynamics may further modulate the both contribution, not only through addition dilution but also new generation of higher-order connected correlations. Although this part is expected to be limited by the decay of fluid kinetic energy, it is crucial to determine the net non-Gaussian signature accessible to the observation. We leave this analysis for future work.

\appendix
\section{Correlation Bounds and Acoustic GW Production}
\label{sec:appendix}

This appendix provides further details of the derivation in Eqs.~(40)--(48). We first establish the spatial properties of connected correlations generated by bubble nucleation and growth. We then examine how these correlations contribute to the acoustic GW spectrum and explain the different time dependence of the Wick and connected terms.

Consider the boolean function $\chi_i$ at a point $\mathbf{x}_i$ at the end of nucleation. We introduce a truncated indicator $\chi_i^{R}$ by excluding bubbles nucleated before $u=-R/v_w$. The truncated indicator depends only on nucleation events within a distance $R$ of $\mathbf{x}_i$. 
For the nucleation rate $\Gamma(u)=\Gamma_0 e^{\beta u}$, the probability of this event is bounded by the expected number of such bubbles,
\begin{equation}
\begin{split}
\Pr(\chi_i\ne\chi_i^{R})
&\leq \frac{4\pi v_w^3\Gamma_0}{3}
       \int_{R/v_w}^{\infty} ds\,s^3e^{-\beta s}\\
&=\Lambda_f e^{-\beta R/v_w}Q_3(z),
\end{split}
\end{equation}
where $\Lambda_f=\dfrac{8\pi v_w^3\Gamma_0}{\beta^4}$ and $Q_3(z)=1+z+z^2/2+z^3/6$. This gives Eq.~(40) and shows that the effect of earlier nucleation events decreases exponentially with $R$.

We next extend this bound to the connected correlation of four points with maximum pairwise separation $D=\max_{i,j}|\mathbf{x}_i-\mathbf{x}_j|$. Choosing $R=D/12$, the four truncated regions cannot all be connected, since a chain of three successive overlaps can span at most $6R<D$. These regions can therefore be divided into at least two independent nonempty groups, implying that their connected four-point correlation vanishes. To obtain a bound for the original indicators, we use the fourth-order cumulant expansion, in which the sum of the absolute values of the coefficients is 26. Replacing one indicator changes the cumulant by at most 26 times its truncation probability. Applying this bound successively to all four indicators gives
\begin{equation}
\left|\langle\chi_1\chi_2\chi_3\chi_4\rangle_c\right|
\leq 104\Lambda_f e^{-\beta D/(12v_w)}
Q_3\!\left(\frac{\beta D}{12v_w}\right),
\end{equation}
which is Eq.~(41). The same bound also holds for points evaluated during the finite period of bubble nucleation and growth. Consequently, the connected conversion correlation is exponentially suppressed at large spatial separations, ensuring its integrability over the relative positions of the four points.

The spatial bound derived above also constrains the acoustic velocity correlations, since the velocity is linearly sourced by the phase-transition indicator $\chi$ over a finite period Eq.~(38). After factoring out the momentum conserving delta function, the Fourier transform of the connected four-point correlation remains bounded, while the smoothing kernel $K_\ell$ suppresses contributions from large acoustic momenta. These properties further constrain the connected contribution Eq.~(43) to the GW spectrum during subsequent free acoustic propagation.

To establish the bound in Eq.~(44), we examine the momentum dependence of the acoustic frequency differences. For $\sigma=-\tau$, the triangle inequality gives $|q-|\mathbf{k}-\mathbf{q}||\leq k$, so the corresponding $\tilde q$ remains bounded at fixed $k$. For $\sigma=\tau$, large frequency differences correspond to large values of $q+|\mathbf{k}-\mathbf{q}|$, whose contributions are suppressed by the smoothing kernel $\widehat K_\ell$. Together with the integrability of the connected four-point correlations and the finite duration of acoustic generation, these properties constrain the two-frequency kernel $H_C(\tilde q,\tilde q';k)$, leading to the bound in Eq.~(44). 

The regularity of $H_C$ allows us to bound the time dependence of the connected contribution in Eq.~(43) using the Fourier factor $F_f$. For the time weights considered here, including the effective weight $f(s)/a(s)$ in an expanding background, their extensions by zero outside the integration interval have total variation at most 2. Integration by parts and Parseval's identity then give
\begin{equation}
|F_f(\delta)|\leq\frac{2}{|\delta|},
\qquad
\int d\delta\,|F_f(\delta)|^2=2\pi E.
\end{equation}

Using these bounds, we split the frequency integral into two regions, $|\delta|\leq 1/E$ and $|\delta|>1/E$. In the first region, the Cauchy--Schwarz inequality gives an upper bound of $2\sqrt{\pi}$, while in the second region, the bound $|F_f(\delta)|\leq 2/|\delta|$ limits the contribution from large frequency differences. Combining both regions yields
\begin{equation}
\int d\delta\,
\frac{|F_f(\delta)|}{1+(\delta/\beta)^2}
\leq J(E),
\end{equation}
where
\begin{equation}
J(E)=2\sqrt{\pi}
+2\ln\!\left[1+(\beta E)^2\right].
\end{equation}
Applying this bound to both frequency integrals in Eq.~(43), we find that the connected contribution during free acoustic propagation grows at most as $J^2(E)$.

The complete connected contribution also includes GWs generated during the finite formation period and their interference with subsequent acoustic emission. The former is bounded independently of $E$, while the interference term involves only one free-acoustic time integral and therefore grows at most as $J(E)$. These additional terms can be absorbed into the overall bound by redefining the $E$-independent coefficient $Q_C(k)$. We thus recover Eq.~(45) for the complete connected contribution.

In contrast, the Wick contribution has a different frequency structure. The Wick contractions identify the acoustic momenta between the two stress factors, yielding the delta function $\delta_D(\tilde q-\tilde q')$ in Eq.~(46). The resonant contribution can therefore be expressed as
\begin{equation}
W_f^{\rm res}(k)
=\int d\delta\,Q_W(\delta;k)|F_f(\delta)|^2.
\end{equation}
For a long period of acoustic propagation in flat spacetime, $|F_f(\delta)|^2/(2\pi E)$ approaches a delta function centered at $\delta=0$. Provided that $Q_W(\delta;k)$ is continuous near $\delta=0$ and $Q_W(0;k)>0$, the leading contribution becomes
\begin{equation}
W_f(k)=2\pi Q_W(0;k)E+o(E),
\end{equation}
as given in Eq.~(47). Contributions from nonresonant terms and the initial formation period do not alter this leading behavior.

Comparing this linear growth with the slower accumulation of the connected contribution yields Eq.~(48). The difference arises because the connected kernel lacks the corresponding delta function and therefore does not exhibit the same linear accumulation. 

\begin{acknowledgments}
This work was supported by the National Natural Science Foundation of China (Grant Nos.12605177),  the National Natural Science Foundation of Henan province(Grant No. 27AQ140008).
\end{acknowledgments}

\bibliographystyle{apsrev4-2}
\bibliography{references}
\end{document}